\documentclass[floatfix,showpacs,aps,twocolumn]{revtex4}
\usepackage{graphicx}
\usepackage{color}
\usepackage{bm}

\newcommand{\figref}[1]{Fig.~\ref{#1}}

\begin{document}
\title{Resistance jumps and the nature of the finite-flux normal phase
in ultra-thin superconducting cylinders}
\author{G.J.~Conduit}
\affiliation{Department of Physics, Ben Gurion University, Beer Sheva 84105, Israel}
\author{Yigal Meir}
\affiliation{Department of Physics, Ben Gurion University, Beer Sheva 84105, Israel}

\date{\today}

\begin{abstract}
Recent observations have revealed the emergence of an unusual normal
phase when a magnetic flux threads an ultra-thin superconducting
cylinder. Moreover, with increasing temperature, the resistance rises
in a series of abrupt jumps. These phenomena are explained using a
novel approach, which allows calculation of the resistance in the
presence of amplitude and phase fluctuations of the superconducting
order parameter, and at the same time introduces a local probe of the
current and chemical potential. It is demonstrated that phase fluctuations
lead to the sequential breakdown of local superconducting phase
correlations, resulting in the formation of normal weak links, which
give rise to the emergence of the normal phase is a stepwise
manner. Finally, specific predictions are made on how the experimental
observations change with the cylinder geometry.
\end{abstract}


\maketitle

Almost half a century ago, Little and Parks~\cite{Little62} performed
one of classic experiments in superconductivity: they demonstrated
that the critical temperature of a cylindrical superconductor (SCR)
varies periodically with the magnetic flux $\Phi$ threading the
cylinder; the period $\Phi_0\equiv hc/2e$ reflects the charge $2e$
of the Cooper pairs. This effect is well understood within the BCS
mean-field model~\cite{Tinkham}, as the kinetic energy of the
electrons depends periodically on the magnetic flux. In fact, it has
been predicted~\cite{deGennes81} and subsequently
observed~\cite{Liu01} that if the cylinder circumference is reduced to
the same order as the SCR coherence length, the flux can drive the SCR
into its normal phase even at the lowest
temperature. Surprisingly~\cite{Liu01}, the resistance of this
low-temperature normal phase, only weakly dependent on temperature, is
considerably smaller than the high temperature, normal state
resistance of that same sample. A subsequent experiment has revealed
even more intriguing results: the resistance increased with rising
temperature in a series of steps~\cite{Wang05}, which broadened with
applied magnetic flux.  The origin of these steps, and the nature of
the resistive state at finite flux have become a subject of much
debate. Originally it was suggested \cite{Wang05} that the steps arise
due to consecutive events of phase separation in the vicinity of a
quantum phase transition (see also \cite{Vafek05}). It was later
demonstrated~\cite{Dao08,Dao09}, however, that such a scenario is
inconsistent with the system parameters, and a spontaneous transition
to a symmetry broken order parameter is not possible.  Alternatively,
a mean-field-like calculation~\cite{Dao08,Dao09} suggested that the
observation might be due to disorder induced fluctuations in the coherence
length. However, it was then claimed~\cite{Wang08} that the large
variation in the coherence length necessary to explain the data was
inconsistent with other features in the experiments. Thus, the
experimental observations remain hitherto unexplained.

In a two-dimensional system at the critical temperature
$T_{\text{KT}}$ phase fluctuations of the pairing amplitude drive
vortices and anti-vortices to unbind and proliferate through the
system~\cite{BKT}. The destruction of global phase coherence drives
the loss of perfect conductance~\cite{Halperin}. It is therefore
imperative for any theory that attempts to describe the loss of
superconductivity in these ultra-thin cylinders to take phase
fluctuations into account. In this letter we utilize a new formulation
\cite{Conduit10} of transport through low-dimensional, possibly
disordered SCRs (attached to two metallic leads), in the presence of
finite temperature and magnetic field, to study a microscopic model of
the Little-Parks effect. This \emph{ab initio} tool, which employs an
exact formula for the current, takes full account of thermal phase
fluctuations of the superconducting (SC) order parameter, while neglecting
its quantum fluctuations. This method has
already been shown~\cite{Conduit10} to reproduce, for example, the
classic Little-Parks effect and the conductance characteristics near
the Kosterlitz-Thouless transition. Moreover, the new tool introduces
a local probe of the normal and SC current and
chemical potential distribution within the sample, giving us the
opportunity to expose and understand the local physical processes that
drive the loss of superconductivity and concomitant steps in the
resistance. In the present case the intermediate region, an ultra-thin
cylinder, sandwiched between two metallic leads, is described by the
disordered negative-$U$ Hubbard lattice model. The parameters that
characterize the model, in addition to temperature $T$ and magnetic
flux $\Phi$, are the on-site attraction $U$ (all energies are
expressed in terms of $t$, the uniform hopping integral), the typical
disorder $W$ -- the width of the distribution of the on-site energies,
and the average density $n$ (see \cite{Conduit10} for more
details). The coupling between the noninteracting leads and the
central region results in a length independent contact resistance,
which can be easily dealt with. The magnetic field is applied parallel to the
axis of the cylinder, which has a finite wall thickness $d$ and inside
radius $r$ (expressed in terms of the lattice constant $a$).

Our strategy to study the Little-Parks effect in an ultra-thin
cylinder is to first establish the geometry and key parameters
required to recover the main experimental phenomena, and then compare
and contrast the observations to experiment. Secondly, we will take
advantage of our new capability to study the microscopic observables
to highlight the underlying mechanisms that drive the emergence of the
normal state.  Finally, we will use our formalism to make predictions
about further observable phenomena that could verify the validity of our approach and 
help pin down the nature of the transition.

\begin{figure}
\includegraphics[width=\linewidth]{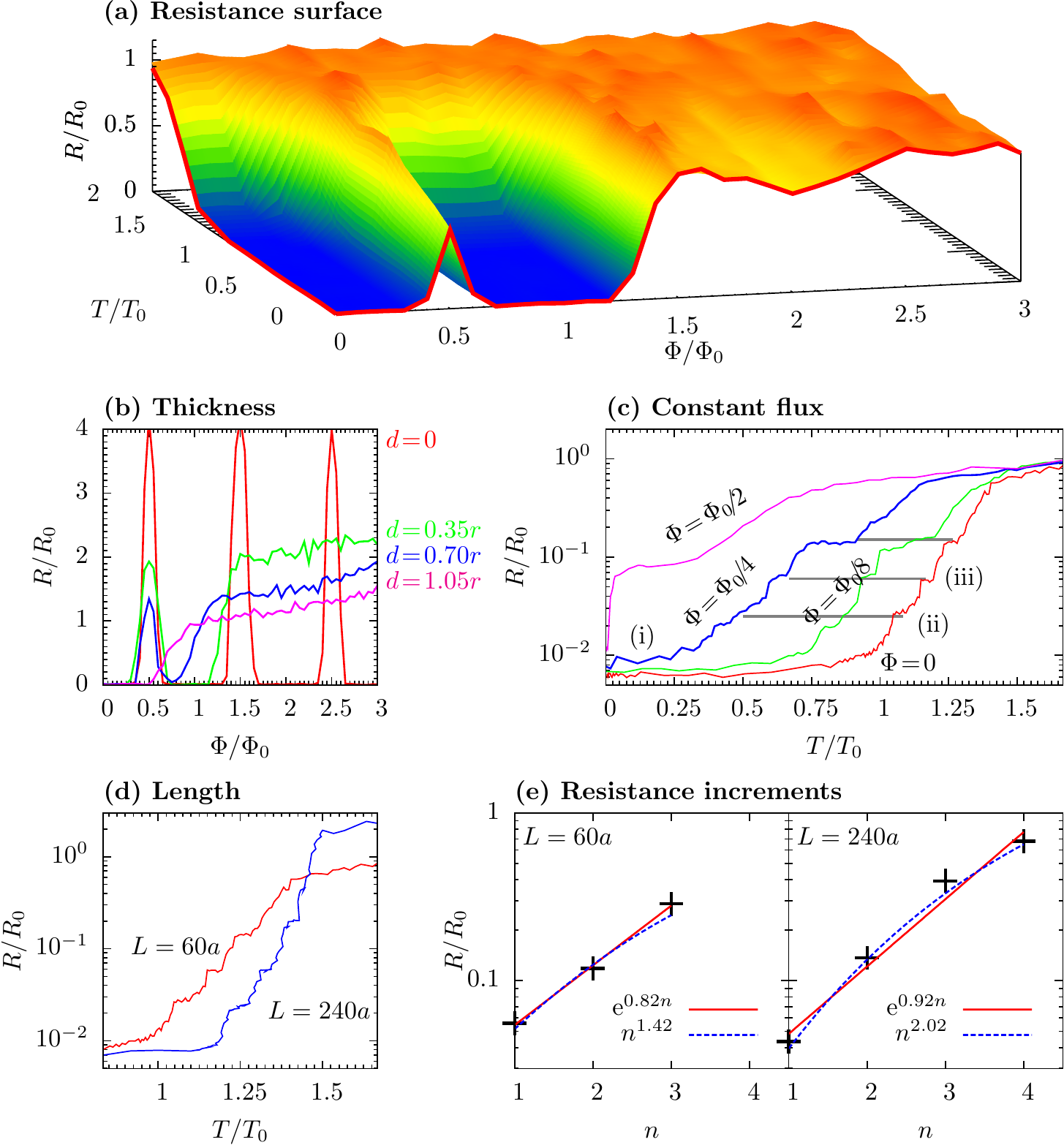}
\caption{(Color online)
(a) The variation of resistance with flux $\Phi$ and temperature $T$,
with the normal state at $\Phi=0$ first emerging at temperature
$T_{0}=0.11t$. $R_{0}=17h/e^2$ is the normal state resistance. The plot
is perfectly symmetric in magnetic flux.
(b) The variation of resistance with flux for four different cylinder
wall thicknesses $d$.
(c) Cuts through the resistance surface at four different magnetic
fluxes. The horizontal gray lines denote the emergence of the first
three steps.
(d) Cuts through the resistance surface at zero flux for
two different length SCRs.
(e) The steps in the resistance against step number for systems of
length $60a$ and $240a$ at $\Phi=0$.}
\label{fig:LPFluxVary}
\end{figure}

In order to determine a suitable geometry for our cylinder, we first
note that the resistance was not a periodic function of the magnetic
field because of flux penetrating through the sample
walls~\cite{Liu01,Wang05}. Our first numerical results in
\figref{fig:LPFluxVary}(b) depict the magnetic field dependence of
the resistance $R$ for several wall thicknesses $d$. We find that the
choice $d\approx0.35r$ (which falls within the typical range of the experimental devices),
has a reasonable resemblance to the experimental results, and the
numerical results presented hereafter are for that choice of thickness
(in all the calculations presented in this Letter, we set the
parameters $U=1.6t$, $W=0.1t$, $n=0.9$, use a cylinder of circumference
$11a$, and length $L=60a$). \figref{fig:LPFluxVary}(a) depicts the
dependence of the resistance with flux and temperature. In addition to
the expected suppression of superconductivity by temperature, it
demonstrates the emergence of the normal phase at low temperature
around $\Phi\approx\Phi_{0}/2$, and the recurrence of the SC phase for
$\Phi_{0}/2<\Phi<\Phi_{0}$. This surface displays the same qualitative
features as the experimental results in Ref.~\cite{Liu01}. In
\figref{fig:LPFluxVary}(c) we plot cuts through the surface at
constant threading flux. The cut taken at zero flux contains a
substantial SC phase region (where the resistance is only due to the
contact resistance to the leads) before the resistance increases into
the normal phase in a series of steps. Increasing flux lowers the
transition temperature, until at half-integer flux the system starts
out in a normal phase whose resistance then increases further with
rising temperature.  This normal state persists to almost $T=0$ at
which point the weakly coupled superconducting (SC) regions become
phase locked and the resistance rapidly drops. This can also possibly
be seen in some of the experimental results~\cite{Wang05}.

Having outlined the main qualitative behavior we now focus more
closely on the emergence of the steps in the resistance, which are
also seen in the experimental results~\cite{Wang05}. For the cuts at
different fluxes in \figref{fig:LPFluxVary}(c), the steps in the
resistance emerge at similar values of the overall resistance roughly
independent of flux. Interestingly, for the longer $L=240a$ wire shown
in \figref{fig:LPFluxVary}(d) more steps emerge. To examine these
steps in more detail,  \figref{fig:LPFluxVary}(e) depicts the
dependence of the resistance step value on the step number, for
cylinders of two different lengths, $L=60a$ and $L=240a$. To reliably
distinguish the steps we focus on only the lowest steps (these steps can be
also positively identified from the successive emergence of boundaries
in the phase coherence plots in the lower panel of
\figref{fig:CurrentMap}, see below). The resistance steps are
consistent with either an exponential (as deduced in the
experimental paper~\cite{Wang05}) or power law dependence.

\begin{figure*}
 \includegraphics[width=1.0\linewidth]{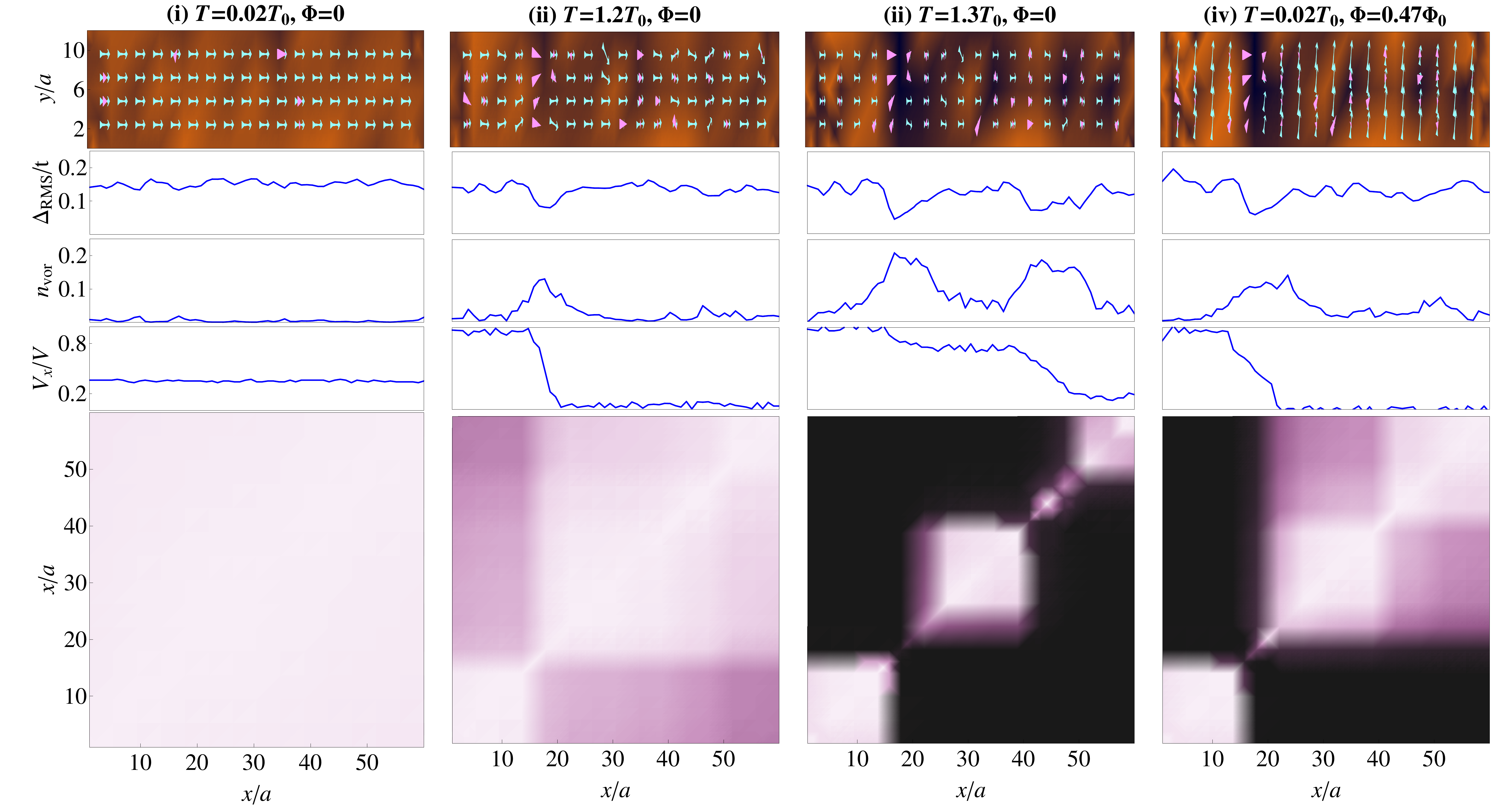} \caption{(Color
 online) The current maps (upper) at four points labeled in
 \figref{fig:LPFluxVary}(e). The normal current is shown by cyan
 pointers and supercurrent in magenta darts, whose length and
 direction reflects the local current flow. The map overlays the
 spatial variation of $|\Delta|$ shown by dark orange. The second row
 of plots show the column averaged variation of $|\Delta|$, the third
 row the column averaged number of vortices per site, and the fourth
 row the local potential. The bottom row shows the column averaged
 correlation $\langle\cos(\theta_{x1}-\theta_{x2})\rangle$, with
 strong correlations in white, and weak in black.}
 \label{fig:CurrentMap}
\end{figure*}

Having reproduced the experimental results, we now take advantage of
our ability to produce local current and voltage maps, and
differentiate between SC (Cooper-pair) and normal current
\cite{Conduit10}, to expose the microscopic mechanism that leads to
the emergence of the normal phase with increasing temperature and
flux. We start at low temperature [point (i) in
\figref{fig:LPFluxVary}(c)] well below $T_{0}$, where the system first
assumes a finite resistance. Here, as we can see in the current map,
[top panel of \figref{fig:CurrentMap}(i)] all the current is SC. This
conclusion is substantiated by the SC order parameter (second panel)
being nearly constant along the wire. Defining the phase of the order
parameter $\theta_{x}$ at a point $x$ along the cylinder averaged over
the circumference, we can see in the lower panel that the long-range
SC correlation function $\langle\cos(\theta_{x1}-\theta_{x2})\rangle$
is constant across the system, demonstrating that the entire SCR is
phase coherent. We also note (third panel) that there are no vortices
in the superconductor, and there is no voltage drop along the sample
(fourth panel), so the only source of resistance is the contact
resistance.

As temperature increases to $T=1.2T_{0}$, [\figref{fig:CurrentMap}(ii),
second panel, corresponding to point (ii) in
\figref{fig:LPFluxVary}(b)], the SC correlations slightly weaken
at around $x/a\simeq17$ for this specific realization of
disorder. While a mean-field calculation would still lead to
long-range coherence in the sample, temperature also enhances phase
fluctuations across the weaker SC region. These weaker local SC
correlations allow nucleation of vortices at lower energy (see
\figref{fig:CurrentMap}(ii) central panel) driving a local loss of
correlation across the junction. This can be seen in the lower panel,
where the system separates into two disconnected strongly
correlated SC regions, giving rise to an
effective Josephson junction, and a finite voltage developing across
the junction (fourth panel). At this temperature the resistance
effectively jumps to a finite value. The current through the system is
naturally reduced, and in fact changes its character from SC to normal
and back to SC as it crosses the junction (top panel).
The bottom panel of \figref{fig:CurrentMap}(ii) indicates also
that weaker correlations are developing between the two sides of
$x/a\simeq45$, which can be thought of as an effective Josephson
junction whose maximal current is still larger than the current in the
system, and thus no voltage drops across it.  As temperature increases
further, there is little change in the resistance, which is basically
due to the normal resistance of the Josephson junction at
$x/a\simeq17$, until thermal fluctuations lead to further nucleation
of vortices, voltage developing across a second Josephson junction at
$x/a\simeq45$, and a second jump in the resistance [see
\figref{fig:CurrentMap}(iii), corresponding to point (iii) in
\figref{fig:LPFluxVary}(c)]. Increasing the temperature further leads 
to the formation of additional weak links, resulting in more jumps in the
resistance.

\begin{figure}
 \includegraphics[width=1\linewidth]{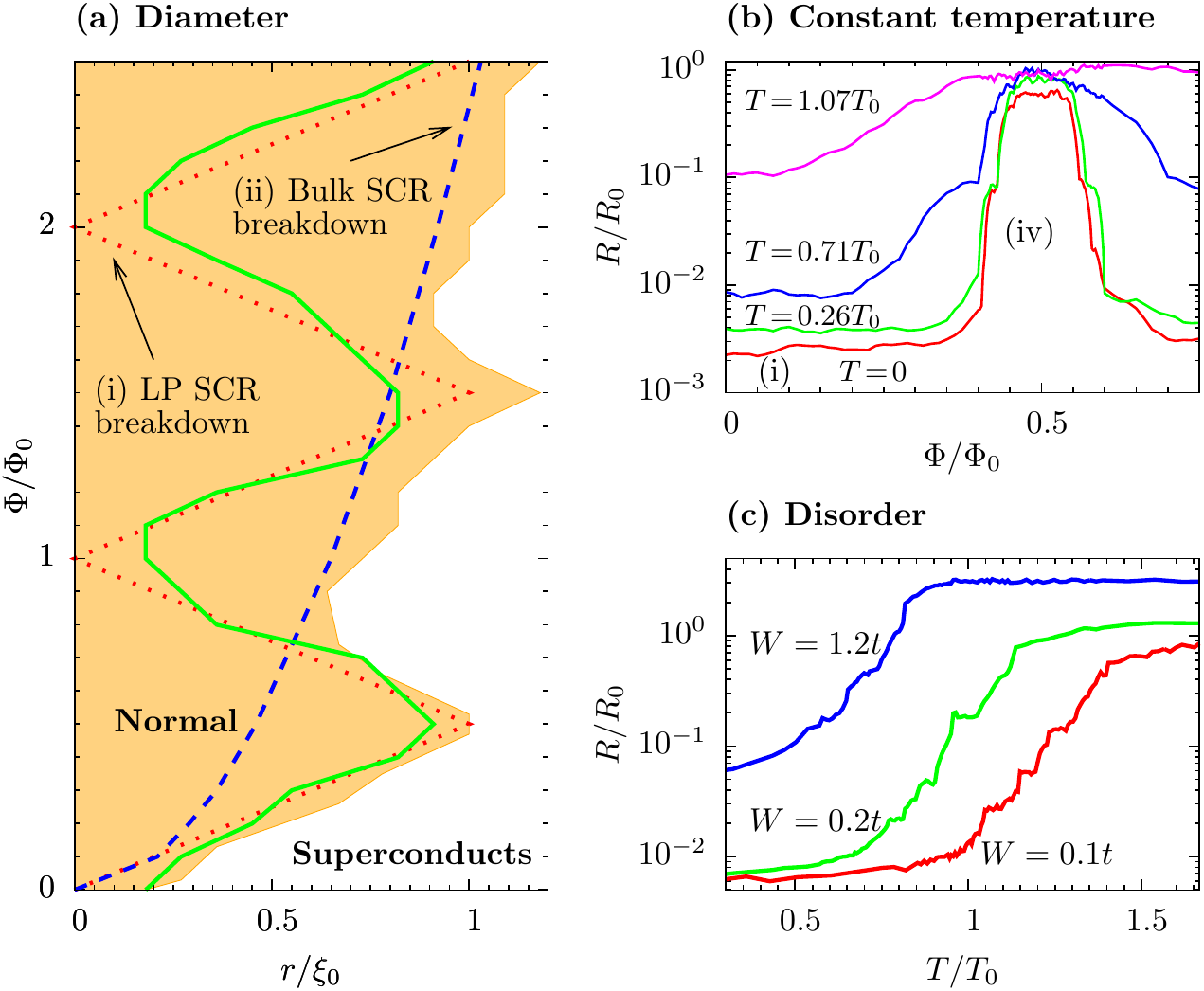} \caption{(Color
 online) The influence of cylinder geometry on the resistance:
 (a) The SCR to normal state transition for different cylinder diameters
 for a $d=0$ cylinder (green solid) and $d=0.35r$ (orange shading).
 The red dotted line (i) shows the mean-field prediction for the LP
 superconductor-insulator transition, and the blue dashed line (ii) shows the
 mean-field prediction for the breakdown of superconductivity due to
 flux penetrating the walls.
 (b) The variation of resistance with flux for four different cylinder
 wall thicknesses $r$.
 (c) The zero flux superconductor-insulator
 transition for three levels of disorder.}
 \label{fig:LittleParksPlot}
\end{figure}

Unlike the mechanism suggested in Refs.\cite{Dao08,Dao09}, one does
not need to invoke large variations in the coherence length of up to
$40\%$ along the sample to quench the SC order. From the data in the
first panel of \figref{fig:CurrentMap}(i) we can estimate the local
coherence length using $\xi=\hbar v_{\text{F}}/\pi\Delta$ as varying
by no more than $\sim15\%$ across the sample. In this low dimensional
system, small variations in the coherence length are sufficient to
induce separation of the system into local SC regions, connected by
normal weak-links, due to the increasing importance of thermal
fluctuations. On the other hand, the system indeed separates into
normal and SC regions, as suggested by Ref.\cite{Wang05}, but not due
to spontaneous phase separation near a critical point, but due to the
interplay of disorder and temperature driven fluctuations.  Similar
behavior is observed upon keeping the temperature fixed, but
increasing the magnetic flux in \figref{fig:CurrentMap}(iv) [see also
point (iv) in \figref{fig:LittleParksPlot}(b)]. Here the loss of
coherence is not driven by fluctuations, but instead the vortices,
induced by the finite magnetic field, disrupt the SC state even at
very low temperatures. Nevertheless, we still expect that as a
function of flux, the resistance will rise in steps, similar to the
steps that emerge with rising temperature. This indeed is borne out by
the numerics, as depicted in
\figref{fig:LittleParksPlot}(b). Furthermore, the lower panel of
\figref{fig:CurrentMap}(iv) shows that for the same system, the
breakdown of the SC phase correlations takes place at the same
location, $x\simeq17a$ as it did with increasing temperature in
\figref{fig:CurrentMap}(iv), so pointing toward the same disorder
driven transition. The observation that the resistance jumps are
almost identical for a longer sample [see \figref{fig:LPFluxVary}(d)],
except that there are more of them, further supports the scenario
where each step is due to the addition of a single normal weak link.

So far we have concentrated on choosing parameters to best fit the
experimental observations. The formalism allows us, naturally, to
explore any range of parameters, and make specific predictions.  We
have already reported in \figref{fig:LPFluxVary}(b) the flux
dependence of the resistance for cylinders of different
thicknesses. As expected, the thicker the cylinder walls, the larger
the magnetic field penetrating the actual sample, and the less
periodic the signal is, until at $d\sim1.05r$, the SC phase does not
reemerge beyond $\Phi=\Phi_0/2$ (the actual value of this thickness
may depend on disorder). To study this in more detail
\figref{fig:LittleParksPlot}(a) depicts the full phase dependence on
wall thickness. For a two-dimensional cylinder ($d=0$) the numerical
separatrix between the SC and normal phase deviates from the
mean-field expectation $r=\xi_{0}\min_{n}|n+\Phi/\Phi_{0}|$ [labeled (i)] at small
cylinder radii $r$, due to the fact that in such quasi one-dimensional
systems, thermal fluctuations are sufficient to disrupt the SC phase
even at zero flux. The filled orange area depicts the SC phase for a
finite-thickness cylinder $d=0.35r$. At small flux we see the same
periodic variation as the $d=0$ cylinder. However, flux penetrating
the walls will itself destroy the SC phase along
\cite{deGennes81} $r=2\xi_0\Phi/\Phi_0\sqrt{1-T/T_{\text{c}}}$ [labeled (ii)],
with $\xi_0$ the SC coherence length used as a fitting parameter,
and so place a bound on the phase boundary.

We have also shown in \figref{fig:LPFluxVary}(d) that the values of
the resistance plateaus depend only weakly on the cylinder length,
though a longer cylinder is predicted to exhibit more
steps. Additionally, \figref{fig:LittleParksPlot}(b) demonstrates that
one should also see steps in the resistance as a function of flux, for
a fixed temperature. Interestingly, these steps persist even above the
critical temperature, which is consistent with the picture presented
above -- the critical temperature corresponds to the formation of the
first normal weak link, but additional normal links are formed with
increasing flux. Finally, in \figref{fig:LittleParksPlot}(c) we plot
the variation of resistance for two higher levels of disorder. At higher
disorder amplitude the transition temperature decreases, and the normal
state resistance increases. Interestingly, the resistance steps persist
even for a highly disordered cylinder $(W=1.2t)$ that exhibits a normal
resistance at low temperature. This prediction could be readily checked
experimentally.

To conclude, we have used \emph{ab initio} simulations to examine
recent experiments on the quantum Little Parks effect. The simulations
demonstrate step-wise destruction of the SC phase, stemming from phase
fluctuations breaking down SC coherence in those parts of the cylinder
that have a weaker BCS order parameter, due to disorder. The formalism
also allowed specific predictions of further phenomena that could
verify this hypothesis.

{\it Acknowledgments:} The authors are grateful to Yuval Oreg for alerting them to this problem.
GJC acknowledges the financial support of the
Royal Commission for the Exhibition of 1851 and the Kreitman
Foundation. This work was also supported by the ISF.


\begin{thebibliography}{99}

\bibitem{Little62}
W.A.~Little and R.D.~Parks, Phys.~Rev.~Lett.~{\bf 9}, 9 (1962).

\bibitem{Tinkham}
M. Tinkham, {\sl Introduction to Superconductivity} (McGraw-Hill, New York, 1996), pp. 128–130.

\bibitem{deGennes81}
P.G.~de Gennes, C.R.~Acad. Sci. (Paris) Ser. II 292, 279 (1981).

\bibitem{Liu01}
Y.~Liu, Y.~Zadorozhny, M.M.~Rosario, B.Y.~Rock, P.T.~Carrigan and H.~Wang,
Science {\bf 294}, 2332 (2001).

\bibitem{Wang05}
H.~Wang, M.M.~Rosario, N.A.~Kurz, B.Y.~Rock, M.~Tian, P.T.~Carrigan and Y.~Liu, Phys.~Rev.~Lett. {\bf 95}, 197003 (2005).

\bibitem{Vafek05}
O.~Vafek, M.R.~Beasley and S.A.~Kivelson, arXiv:0505.688 (2005).

\bibitem{Dao08}
V.H.~Dao and L.F.~Chibotaru, Phys.~Rev.~Lett. {\bf 101}, 229701 (2008).

\bibitem{Dao09}
V.H.~Dao and L.F.~Chibotaru, Phys.~Rev.~B {\bf 79}, 134524 (2009).

\bibitem{Wang08}
H.~Wang, M.M.~Rosario, N.A.~Kurz, B.Y.~Rock, M.~Tian, P.T.~Carrigan and Y.~Liu, Phys.~Rev.~Lett. {\bf 101}, 229702 (2008).

\bibitem{BKT}
V.L.~Berezinskii, Sov. Phys. JETP {\bf 32}, 493 (1971);
J.M.~Kosterlitz and D.J.~Thouless, Journal of Physics C: Solid State Physics, {\bf 6}, 1181, (1973).

\bibitem{Halperin}
For a review, see B.I.~Halperin, G.~Refael and E.~Demler, arXiv:1005.3347.

\bibitem{Conduit10}
G.J.~Conduit and Y.~Meir, arXiv:1102.1604.

%\bibitem{Meyers71}
%L.~Meyers and R.~Meservey, Phys.~Rev.~B {\bf 4}, 824 (1971).

\end{thebibliography}
\end{document}